\documentclass[11pt]{article}

\usepackage{amsmath,amssymb,amsfonts,latexsym,amscd,amsthm}
\usepackage{fancyhdr,color,epsfig,setspace,caption,times}
\usepackage[colorlinks=true,plainpages=false,linktocpage,linkcolor=blue,citecolor=green,pagebackref]{hyperref}
\usepackage[letterpaper,left=1.5in,top=1in,bottom=1in,right=1.5in,includeheadfoot,headheight=16.6pt]{geometry}
\usepackage{graphicx}
\newtheorem{definition}{Definition}[subsection]
\usepackage{longtable}
\usepackage{bbm}
\usepackage{dsfont}
\def\ket#1{\mathinner{|{#1}\rangle}}

\begin{document}

\title{Properly Quantized History Dependent Parrondo Games, Markov Processes, and Multiplexing Circuits}
\author{Steven A. Bleiler\footnote{Portland State University, Portland, Oregon 97207-0751.}, \hspace{.1in} Faisal Shah Khan\footnote{Portland State University, Portland, Oregon 97207-0751. Email: faisal@pdx.edu}}
\maketitle

\begin{abstract}
In the context of quantum information theory, ``quantization'' of various mathematical and computational constructions is said to occur upon the replacement, at various points in the construction, of the classical randomization notion of probability distribution with higher order randomization notions from quantum mechanics such as quantum superposition with measurement.  For this to be done ``properly'', a faithful copy of the original construction is required to exist within the new ``quantum'' one, just as is required when a function is extended to a larger domain. Here procedures for extending history dependent Parrondo games, Markov processes and multiplexing circuits to their "quantum" versions are analyzed from a game theoretic viewpoint, and from this viewpoint, proper quantizations developed.
\end{abstract}

\section{Introduction}

For the most part, mathematicians view games as functions, a point of view that allows the enlargement of the sets of possible strategies, outcomes and solutions in a game without necessarily eliminating the players abilities to play the original game in the new context. One way this is achieved is by identifying a game with its payoff function and then extending this function's domain. Since an extended domain necessarily restricts to the original one, the original game can be recovered from the new ``extended'' game when appropriate restrictions are introduced. This allows meaningful comparison between the game theoretic properties of the two versions of the game. The use of domain extension is ubiquitous in game theory and is most commonly recognized in the form of {\it mixed strategies}, that is, randomizations between the so-called {\it pure} strategic choices of a player. 

To elaborate, recall that a key goal in the study of multi-player, non-cooperative games is the identification of potential Nash equilibria. Informally, a Nash equilibrium occurs when each player chooses to play a strategy that is a best reply to the choice of strategies of all the other players. In other words, unilateral deviation from the choice of strategy at a Nash equilibrium by any player cannot improve that player's payoff in the game. However, Nash equilibria need not be optimal and in other cases they may not even exist. In such situations, games are frequently ``enlarged'' via the definition of an extended set of strategic choices and an analysis of the extended game performed. As an example for finite games, passing to mixed strategies often gives rise to Nash equilibria in the ``mixed game'' that simply do not exist in the original game.  Formally, the mixed game results from an extension of the domain of the payoff function to include randomization between the pure strategies in the form of probability distributions over the pure strategies. That mixed strategies arise from domain extension is also clear from the fact that a faithful copy of every pure strategy set sits inside the corresponding mixed strategy set by considering a pure strategy as  played with certainty.  The so-called mixed strategy equilibria sit outside the collection of pure strategies within the mixed ones, and are thus considered as ``new'' equilibria of the original game.

About a decade ago, Meyer \cite{Meyer} proposed the extension of the domain of a game's payoff function so as to include quantum mechanical operations. A concrete example of such an extension was provided soon after by Eisert, Wilkens, and Lewenstein \cite{Eisert}, and applied to the game Prisoner's Dilemma. The area of study arising from these ideas has come to be known as {\it quantum} game theory. Typically, research in the subject looks for different than usual behavior of the payoff function of an $n$ player game under quantization, that is as mentioned above the replacement at various points in the payoff functions definition of probability distribution by quantum superposition and measurement.  This typically involves the replacement of strategic choices or of a family of outcomes by qudits, that is quantum systems having $d$ "pure" quantum states.  Also typically, quantum operations on each qudit are then considered as a set of {\it quantum strategies} for the players. The different than usual behavior is often the occurrence of Nash equilibria that were unavailable in the original game. Following these heuristics produces a {\it quantized game} which is referred to as a {\it quantization} of the original game. 

Because of the lack of explicit reference to any mathematically formal approach of domain extension, these heuristics sometimes produce quantizations that are not true extensions. In such cases, it is impossible to meaningfully compare any game-theoretic results that these quantization generate, such as Nash equilibria, with the results from the original game. Indeed, such quantizations truly ''change the game''. On the other hand, {\it proper} quantizations are true extensions and necessarily restrict to the original game, making possible meaningful comparison between the results of the original game and the quantized one. A formal approach to game quantization via generalizing mixtures developed by one of the authors \cite{Bleiler} is utilized herein to develop proper quantizations of history dependent Parrondo games.

It should be noted that games are not the only informatic or computational constructions and processes currently undergoing extension and analysis via quantization. Two prominent areas of study are Markov processes \cite{Kummerer} and the so-called {\it multiplexing circuits} \cite{FSK:06}. Concerns regarding the existence of faithfully embedded versions of the classical object within the quantized one have also arisen in these areas.

The problems here are also more subtle, because in these areas it is stochasticity, as opposed to probability distribution, serving as the classical component of the construction being replaced by quantum mechanical operations. In particular, the frequently considered replacement quantum concept of completely positive operators and measurement for Markov processes \cite{Kummerer} is also more general than just (normalized) quantum superposition and measurement, forming what in game theoretic language might be termed mixed quantum superposition, i.e. a non-trivial probability distribution over the collection of superpositions. By jumping to this most general form of quantum probability, the existence issue of faithfully embedded copies of the classical process becomes clouded. As illustrated here, and discussed in more detail in a subsequent publication \cite{BK2}, clarity on this issue is gained by initially restricting consideration to quantum Markov processes obtained through the replacement of stochasticity by (normalized) quantum superposition and measurement, and subsequently following to the more general situation.
 
As for quantum multiplexing circuits, motivated by the similarity between the informational behavior of classical multiplexing circuits and certain quantum logic circuits, Shende et al coined the term {\it quantum multiplexer} for the latter in \cite{shende:05}. To be precise, some of the bits in a multiplexing circuit are acted upon by appropriate logic gates under the control of the logical values of some other bits in the circuit. Quantum circuits exhibiting a similar structure, such as the one for the controlled-NOT gate, are also considered under this formulation as quantum multiplexers. However, this definition of a quantum multiplexer is far too informal, allowing for the possibility that a given classical multiplexer may be identified with a whole class of distinct quantum multiplexers. Thus the relation between a multiplexer and its "quantizations" is not functional but relational, and the question of preservation of a faithful copy of the original multiplexer in the quantized one becomes ill defined. A functional relation between the original and quantum versions of a particular multiplexer that arises in the context of history dependent Parrondo games is established in section \ref{prop quantum A and B'}, that at the same time establishes the notion of {\it proper} quantization for multiplexers.  In particular within the quantum versions of the multiplexer lies faithfully embedded copies of the classical to which the quantum multiplexer could be restricted.

For the Markov processes and multiplexing circuits considered here, the focus of attention on (normalized) quantum superposition and measurement also allows the successful resolution of the question of what constitutes an appropriate evaluative quantum analogue of the stable state of a Markov process or as expressed in the context of quantum multiplexers what constitutes the appropriative evaluative initial state. As mentioned above, further discussion of the issues expressed in the previous paragraphs and an answer to the initial state question for the more general contexts appears in a subsequent publication \cite{BK2}.

%Need some references here%

\section{History Dependent Parrondo Games}\label{HD}

Parrondo et. al first formulated such games in \cite{Parrondo}. The subject of Parrondo games has seen much research activity since then. Parrondo games typically involve the flipping of biased coins and yield only expected payoffs. A Parrondo game whose expected payoff is positive is said to be {\it winning}. If the expected payoff is negative, the game is said to be {\it losing}, and if the expected payoff is $0$, the game is said to be {\it fair}. 

Parrondo games are of interest because sequences of such games occasionally exhibit the \emph{Parrondo effect}; that is, when two or more losing games are appropriately sequenced, the resulting combined game is winning. Frequently, this sequence is {\it randomized} which means that the game played at each stage of the sequence is chosen at random with respect to a particular probability distribution over the games being sequenced. A comprehensive survey of Parrondo games and the Parrondo effect by Harmer and Abbott can be found in \cite{Harmer}.  

A special type of Parrondo games is the history dependent Parrondo game, introduced in \cite{Parrondo} by Parrondo et al. This game is again a biased coin flipping game, where now the choice of the biased coin depends on the history of the game thus far, as opposed to the modular value of the capital. A history dependent Parrondo game $B'$ with a two stage history is reproduced in Table \ref{table:gameB HD}.

As above, let $X(t)$ be the capital available to the player at time $t$. At stage $t$, this capital goes up or down by one unit, the probability of gain determined by the biased coin used at that stage. Obtain a Markov process by setting  
\begin{equation}\label{Markov process}
Y(t)=\left(\begin{array}{c}
X(t)-X(t-1) \\ X(t-1)-X(t-2)
\end{array}\right).
\end{equation}
\begin{table}
\begin{center}
\begin{tabular}{c|c|c|c|c}
 Before last & Last& Coin & Prob. of gain & Prob. of loss \\
$t-2$ & $t-1$& ~ & at $t$ & at $t$ \\\hline
gain & gain & $B_1'$ & $p_1$ & $1-p_1$\\ gain & loss & $B_2'$ & $p_2$
& $1-p_2$ \\ loss & gain & $B_3'$ & $p_3$ & $1-p_3$ \\ loss & loss &
$B_4'$ & $p_4$ & $1-p_4$ \\
\end{tabular}
\end{center}
\caption{History dependent game $B'$.} 
\label{table:gameB HD}
\end{table}
This allows one to analyze the long term behavior of the capital in game $B'$ via the stationary state of the process $Y(t)$. The transition matrix for this process is 
\begin{equation}\label{transition matrix}
X=\left(\begin{array}{cccc}
p_1 & 0 & p_3 & 0 \\  1-p_1 & 0 & 1-p_3 & 0 \\ 0 & p_2 & 0 & p_4 \\ 0 & 1-p_2 & 0 & 1-p_4 
\end{array}\right)
\end{equation}
The stationary state can be computed from the following equations
$$
p_1\pi_1+p_3\pi_3=\pi_1
$$
$$
(1-p_1)\pi_1+(1-p_3)\pi_3=\pi_2
$$
$$
p_2\pi_2+p_4\pi_4=\pi_3
$$
$$
(1-p_2)\pi_2+(1-p_4)\pi_4=\pi_4
$$
and is given by 
\begin{equation}\label{stationary state}
s=\left(\begin{array}{c}
\pi_1 \\ \pi_2 \\\pi_3 \\\pi_4 
\end{array}\right)=\frac{1}{N}\left(\begin{array}{c}
p_3p_4  \\ p_4(1-p_1) \\ p_4(1-p_1)  \\ (1-p_1)(1-p_2)
\end{array}\right)
\end{equation}
after setting the free variable $v_4=(1-p_1)(1-p_2)$ and normalization constant
$$
N=\sqrt{\sum^4_{j=1}(\pi_j)^2}=\sqrt{(p_3p_4)^2+2\left[(1-p_1)p_4\right]^2+\left[(1-p_1)(1-p_2)\right]^2}
$$
which simplifies to 
$$ 
N=(1-p_1)(2p_4+1-p_2)+p_3p_4.
$$ 

Consequently, the probability of gain in a generic run of the game $B'$ is 
\begin{equation}\label{p of win B'}
p^{B'}_{\rm gain}=\frac{1}{N}\sum_{j=1}^{4}\pi_{j}p_{j}=\frac{p_4\left(p_3+1-p_1\right)}{\left(1-p_1\right)\left(2p_4+1-p_2\right)+p_3p_4}
\end{equation}
\noindent where $\pi_{j}$ is the probability that a certain history $j$, represented in binary format, will occur, while $p_{j}$ is the probability of gain upon the flip of the last coin corresponding to history $j$. The expression for $p^{B'}_{\rm gain}$ simplifies to  
\begin{equation}\label{p win class}
p^{B'}_{\rm gain} =1/(2+x/y)
\end{equation}
with
\begin{equation}\label{s condition}
y=p_4(p_3+1-p_1)>0
\end{equation}
for any choice of the probabilities $p_1, \dots p_4$, and
\begin{equation}\label{c condition}
x=(1-p_1)(1-p_2)-p_3p_4.
\end{equation}
Therefore, game $B'$ obeys the following
rule: if $x < 0$, $B'$ is winning, that is, has positive expected payoff; if $x= 0$, $B'$ is
fair; and if $x> 0$, $B'$ is losing, that is, has negative expected payoff. 

%In section \ref{QHDPG}, a concise formal description of the history dependent Parrondo games will be given so as to put them in a domain extension context. 

\subsection{Randomized Combinations of History Dependent Parrondo Games}\label{rand A and B'}
Consider now the two stage history dependent game obtained by randomly sequencing the games $B'$ and $B''$ where each of $B'$ and $B''$ are history dependent Parrondo games with two stage histories. This can be formally considered as a real convex linear combination of the games $B'$ and $B''$, where the coefficients on $B'$ and $B''$ are given by $r$, the probability that the game $B'$ is played at a given stage, and $(1-r)$, the probability that the game $B''$ is played at a given stage. This is because the transition matrix of the Markov process associated to the randomized sequence is obtained from the transition matrices $T'$ and $T''$ for the games $B'$ and $B''$, respectively, by taking the real convex combination $rT'+(1-r)T''$. Explicitly, let 
\begin{equation}\label{T' matrix}
T'=\left(\begin{array}{cccc}
\alpha_1 & 0 & \alpha_3 & 0 \\  1-\alpha_1 & 0 & 1-\alpha_3 & 0 \\ 0 & \alpha_2 & 0 & \alpha_4 \\ 0 & 1-\alpha_2 & 0 & 1-\alpha_4 
\end{array}\right)
\end{equation} 
and
\begin{equation} \label{T'' matrix}
T''=\left(\begin{array}{cccc}
\beta_1 & 0 & \beta_3 & 0 \\  1-\beta_1 & 0 & 1-\beta_3 & 0 \\ 0 & \beta_2 & 0 & \beta_4 \\ 0 & 1-\beta_2 & 0 & 1-\beta_4 
\end{array}\right).
\end{equation} 
with $\alpha_j,\beta_j \in [0,1]$ representing the probability of gain for the $j$ coin in games $B'$ and $B''$ respectively. Then the transition matrix $rT'+(1-r)T''$ of the Markov process for the randomized sequence of $B'$ and $B''$ consists of entries $t_j=r\alpha_j+(1-r)(\beta_j)$ and $1-t_j=r(1-\alpha_j)+(1-r)(1-\beta_j)$ in the appropriate locations. Call this randomized sequence of games $B'$ and $B''$ the history dependent game $B'B''$ with probability of gain $t_j$. The stable state, computed in exactly the same fashion as the stable state for the game $B'$ in section \ref{HD} above, has form
\begin{equation}\label{stat state mix general}
\tau=\left(\begin{array}{c}
\tau_1 \\ \tau_2 \\ \tau_3 \\ \tau_4 
\end{array}\right)=\frac{1}{R}\left(\begin{array}{c}
t_3t_4 \\ t_4(1-t_1) \\ t_4(1-t_1) \\ (1-t_1)(1-t_2)
\end{array}\right)
\end{equation}
with $R=\sum^{4}_{j=1}\tau_j$ a normalization constant. Using the stable state, the probability of gain in the game $B'B''$ is computed to be 
\begin{equation}\label{p of gain in B'B''}
p^{B'B''}_{\rm gain}=\frac{1}{R}\sum_{j=1}^{4}\tau_{j}t_{j}=\frac{t_4\left(t_3+1-t_1\right)}{\left(1-t_1\right)\left(2t_4+1-t_2\right)+t_3t_4}.
\end{equation}
Just as in case of the game $B'$, the expression for $p^{B'B''}_{\rm gain}$ reduces to  
\begin{equation}\label{p win class B' and B''}
p^{B'B''}_{\rm gain} =1/(2+x'/y')
\end{equation}
with 
\begin{equation}\label{s condition B'B''}
y'=t_4(t_3+1-t_1)>0
\end{equation}
for any choice of the probabilities $t_1, \dots t_4$, and
\begin{equation}\label{c condition B'B''}
x'=(1-t_1)(1-t_2)-t_3t_4.
\end{equation}
The game $B'B''$ therefore behaves entirely like the game $B'$, following the 
rule: if $x' < 0$, $B'B''$ is winning, that is, has positive expected payoff; if $x'= 0$, $B'B''$ is
fair; and $x'> 0$, $B'B''$ is losing, that is, has negative expected payoff.

It is therefore possible to adjust the values of the $\alpha_j$ and $\beta_j$ in games $B'$ and $B''$ so that they are individually losing, but the combined game $B'B''$ is now winning. This is the Parrondo effect. In the present example, the Parrondo effect occurs when 
\begin{equation}\label{D}
(1-\alpha_3)(1-\alpha_4) > \alpha_1\alpha_2
\end{equation}
\begin{equation}\label{E}
(1-\beta_3)(1-\beta_4)>\beta_1\beta_2
\end{equation}
and
\begin{equation}\label{F}
(1-t_3)(1-t_4)<t_1t_2.
\end{equation}
The reader is referred to \cite{Kay:03} for a detailed analysis of the values of the parameters which lead to the Parrondo effect in such games.

Restricting to the original work of Parrondo et al, a special case occurs when we consider one of the games in the randomized sequence to be of type $A$. That is, flipping a single biased coin which on the surface appears to have no history dependence. However, note that such a game may be interpreted as a history dependent Parrondo game with a two stage history where the coin used in $A$ is employed for every history. Call such a history dependent game $A'$. The transition matrix for $A'$ takes the form
\begin{equation}\label{tran mat A'}
\Delta=\left(\begin{array}{cccc}
p & 0 & p & 0 \\  1-p & 0 & 1-p & 0 \\ 0 & p& 0 & p\\ 0 & 1-p& 0 & 1-p
\end{array}\right).
\end{equation} 
%and the stable state is 
%\begin{equation}\label{stable state A}
%y=\left(\begin{array}{c}
%y_1 \\ y_2 \\ y_3 \\ y_4 
%\end{array}\right)=\frac{1}{N}\left(\begin{array}{c}
%p^2  \\ p(1-p) \\ p(1-p)  \\ (1-p)^2
%\end{array}\right)
%\end{equation}
Now, forming randomized sequences of games $A'$ and $B'$ is seen to agree with the forming of convex linear combinations mentioned above. In particular, as analyzed in \cite{Parrondo} if games $A'$ and $B'$ are now sequenced randomly with equal probability, the Markov process for the randomized sequence is given with transition matrix containing the entries $q_j=\frac{1}{2}(\alpha_j+p)$ and $1-q_j=\frac{1}{2}[(1-\alpha_j)+(1-p)]$ in the appropriate locations (recall that the probability of win for game $A$ is $p$), and has stationary state
\begin{equation}\label{stat state mix}
\rho=\left(\begin{array}{c}
\rho_1 \\ \rho_2 \\ \rho_3 \\ \rho_4 
\end{array}\right)=\frac{1}{M}\left(\begin{array}{c}
q_3q_4 \\ q_4(1-q_1) \\ q_4(1-q_1) \\ (1-q_1)(1-q_2)
\end{array}\right)
\end{equation}
Denote this randomized sequence of games $A'$ and $B'$ by $A'B'$.
The probability of gain in the game $A'B'$ is
\begin{equation}\label{p of win in AB'}
p^{A'B'}_{\rm gain}=\frac{1}{M}\sum_{j=1}^{4}\rho_{j}q_{j}=\frac{q_4\left(q_3+1-q_1\right)}{\left(1-q_1\right)\left(2q_4+1-q_2\right)+q_3q_4}
\end{equation}
As in the more general case of the game $B'B''$, it is now possible to adjust the values of the parameters $p$ and $p_j$'s in games $A'$ and $B'$ so that they are individually losing, but the combined game $A'B'$ is now winning. This happens when 
\begin{equation}\label{A}
1-p > p
\end{equation}
\begin{equation}\label{B}
(1-\alpha_3)(1-\alpha_4)>\alpha_1\alpha_2
\end{equation}
and
\begin{equation}\label{C}
(1-q_3)(1-q_4)<q_1q_2.
\end{equation}
Parrondo et al show in \cite{Parrondo} that when $p=\frac{1}{2}-\epsilon$, $\alpha_1=\frac{9}{10}-\epsilon$, $\alpha_2=\alpha_3=\frac{1}{4}-\epsilon$, $\alpha_4=\frac{7}{10}-\epsilon$, and $\epsilon <\frac{1}{168}$, the inequalities (\ref{A})-(\ref{C}) are satisfied. This is Parrondo et al's original example of the Parrondo effect for history dependent Parrondo games. 

Next we review pertinent features of the formal approach to games developed by Bleiler that puts game quantization in the context of domain extension.

\section{A Formal Approach to Games}\label{Bformal}

We start with a formal definition.
\begin{definition}{\rm  
Given a set $\{ 1, 2, \cdots, n \}$ of players, for each player a set $S_i$ $(i=1, \cdots, n)$ of so-called \emph{pure strategies}, and a set $\Omega_i$ $(i=1, \cdots, n)$ of \emph{possible outcomes}, a \emph{normal form game} $G$ is a vector-valued function whose domain is the Cartesian product of the $S_i$'s and whose range is the Cartesian product of the $\Omega_i$'s. In symbols 
$$
G: \prod_{i=1}^n S_i \longrightarrow \prod_{i=1}^n \Omega_i                      
$$
The function $G$ is referred to as the \emph{payoff function}.}
\end{definition}

Here a \emph{play} of the game is a choice by each player of a particular strategy $s_i$ the collection of which forms a \emph{strategy profile} $(s_1, \cdots, s_n)$ whose corresponding \emph{outcome profile} is $G(s_1, \cdots, s_n)=(\omega_1, \cdots, \omega_n)$, where the $\omega_i$'s represent each player's individual outcome. Note that by assigning a real valued \emph{utility} to each player which quantifies that player's preferences over the various outcomes, we can without loss of generality, assume that the $\Omega_i$'s are all copies of $\mathbb{R}$, the field of real numbers.

In game theory, a {\it rational} players' concern is the identification of a strategy that guarantees a maximal utility. %As this is not usually possible, a \emph{security strategy}, that is, a strategic choice that guarantees an explicit lower bound to the utility received, is also sought. 
For a fixed $(n-1)$-tuple of opponents' strategies then, rational players seek a \emph{best reply}, that is a strategy $s^*$ that delivers a utility at least as great, if not greater, than any other strategy $s$. When every player can identify such a strategy, the resulting strategy profile is called a {\it Nash equilibrium} or occasionally just an {\it equilibrium} of the normal form game $G$. Other ways of expressing this concept include the observation that no player can increase his or her payoffs by unilaterally deviating from his or her equilibrium strategy, or that at equilibrium all of a player's opponents are indifferent to that player's strategic choice.

However, normal form games need not have Nash equilibria amongst the pure strategy profiles. As remarked above, game theoretic formalism now calls upon the theorist to extend the normal form game $G$ by enlarging the domain and extending the payoff function. Of course, the question of if and how a given function extends is a time honored problem in mathematics and the careful application of the mathematics of extension is what will drive the formalism for quantization. In the classical theory, the standard extension at this point is constructed by allowing each player to randomize between his strategic choices, a process referred to as {\it mixing}.

\subsection{Randomization as Domain Extension}

A \emph{mixed strategy} for player $i$ is an element of the set of probability distributions over the set of pure strategies $S_i$. Formally, for a given set $X$, denote the probability distributions over $X$ by $\Delta(X)$ and note that when $X$ is finite, with $k$ elements say, the set $\Delta(X)$ is just the $k-1$ dimensional simplex $\Delta^{(k-1)}$ over $X$, i.e., the set of real convex linear combinations of elements of $X$. Of course, we can embed $X$ into $\Delta(X)$ by considering the element $x$ as mapped to the probability distribution which assigns 1 to $x$ and 0 to everything else. For a given game $G$, denote this embedding of $S_i$ into $\Delta(S_i)$ by $e_i$.

Let $p=\left(p_1, \dots, p_n\right)$ be a mixed strategy profile. Then $p$ induces the \emph{product distribution} over the product $\prod S_i$. Taking the push out by $G$ of the product distribution (i.e., given a probability distribution over strategy profiles, replace the profiles with their images under $G$) then gives a probability distribution over the image of $G$, ${\rm Im}G$, considered as a multi-set. Following this by the expectation operator $E$, we obtain the {\it expected outcome} of the profile $p$. Now our game $G$ can be extended to a new, larger game $G^{mix}$.

\begin{definition}{\rm 
 Assigning the expected outcome to each mixed strategy profile we obtain the extended game 
$$
G^{mix}: \prod \Delta(S_i) \rightarrow \prod \Omega_i
$$}
\end{definition}

\begin{figure}
\centerline{\includegraphics[scale=0.22]{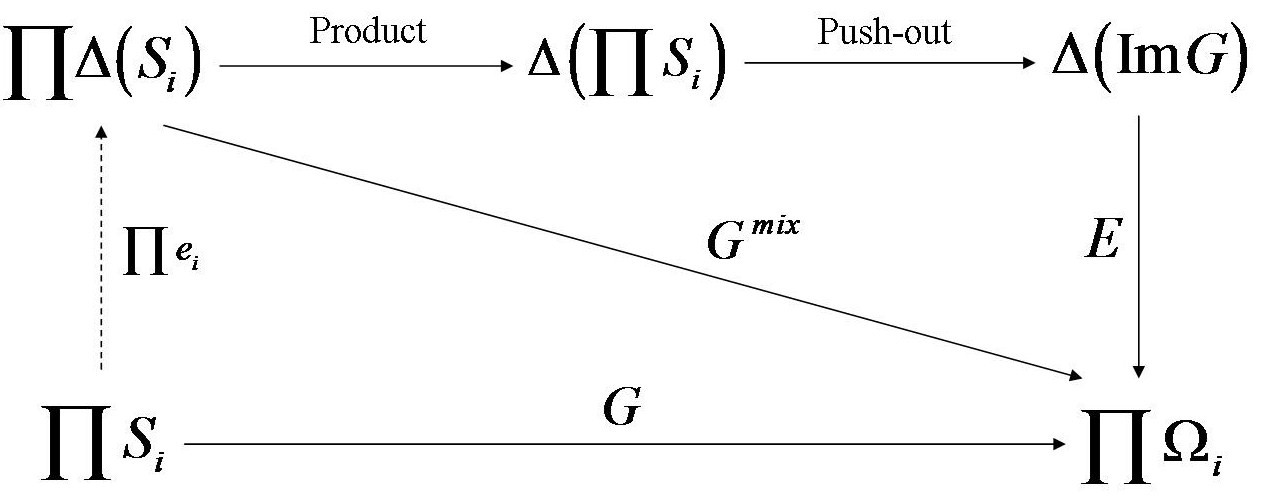}}
\caption{\small{Extension of the game $G$ to $G^{mix}$.}}
\label{Gmix}
\end{figure}

Note $G^{mix}$ is a true extension of $G$ as $G^{mix} \circ \Pi e_i = G$; that is, the diagram in Figure \ref{Gmix} is commutative.

%As remarked above, Nash's famous theorem  \cite{Nash} says that if the $S_i$ are all finite, then there always exists an equilibrium in $G^{mix}$. Unfortunately, this equilibrium is called a \emph{mixed strategy equilibrium for $G$}, when it is not an equilibrium of $G$ at all, the abusive terminology confusing $G$ with its image, Im$G$. 
%%
Having placed a game $G$ and the corresponding game $G^{mix}$ in the domain extension context, the next natural step is to place the notions of mediated communication and {\it correlated} equilibrium \cite{Aumann, Myerson} in a similar context. However, since the latter have no direct relevance tot he topic of this article, we simply refer the reader to \cite{Bleiler} for details.

\subsection{Quantization As Domain Extension}

%The Bleiler formalism asserts that some of the controversies surrounding quantum game theory may be resolved if one focuses on the quantization of the {\it payoffs} of the original game $G$, and expresses the quantized version of $G$ as a (proper) extension of the original payout function in the set-theoretic sense, just as in the classical case. 
Classically, probability distributions over the outcomes of a game $G$ (the image of $G$) were constructed. Now the goal is to pass to a more general notion of randomization, that of quantum superposition. Begin then with a Hilbert space $\mathcal{H}$ that is a complex vector space equipped with an inner product. For the purpose here assume that $\mathcal{H}$ is finite dimensional, and that there exists a finite set $X$ which is in one-to-one correspondence with an orthogonal basis $\mathcal{B}$ of $\mathcal{H}$. 
%\begin{definition}\label{superposition}{\rm
%By a {\it quantum superposition} of elements of $X$ with respect to the basis $\mathcal{B}$ we mean a complex projective linear combination of elements of $X$; that is, a representative of an equivalence class of complex linear combinations where the equivalence between combinations is given by non-zero scalar multiplication.}
%\end{definition}
%
%Quantum mechanics calls this scalar a {\it phase}.
When the context is clear as to the basis to which the set $X$ is identified, denote the set of quantum superpositions for $X$ as $\mathcal{Q}S(X)$. Of course, it is also possible to define quantum superpositions for infinite sets, but for the purpose here, one need not be so general. What follows can be easily generalized to the infinite case. 

As mentioned above, the underlying space of complex linear combinations is a Hilbert space; therefore, we can assign a length to each quantum superposition and, up to phase, always represent a given quantum superposition by another that has length~1. %This process is called {\it normalization} and is frequently useful. 

For each quantum superposition of $X$ we can obtain a probability distribution over $X$ by assigning to each component the ratio of the square of the length of its coefficient to the square of the length of the combination. This assignment is in fact functional, and is abusively referred to as measurement. Formally:

\begin{definition}{\rm 
{\it Quantum measurement with respect to $X$} is the function 
$$
q^{meas}_X: \mathcal{Q}S(X) \longrightarrow  \Delta(X)
$$
given by 
$$
\alpha x + \beta y \longmapsto \left(\frac{\left|\alpha\right|^2}{\left|\alpha\right|^2+\left|\beta\right|^2}, \frac{\left|\beta\right|^2}{\left|\alpha\right|^2+\left|\beta\right|^2}\right)
$$
}
\end{definition}
Note that geometrically, quantum measurement is defined by projecting a normalized quantum superposition onto the various elements of the normalized basis $\mathcal{B}$. Denote quantum measurement by $q^{meas}$ if the set $X$ is clear from the context.

Now given a finite $n$-player game $G$, suppose we have a collection $\mathcal{Q}_1, \dots, \mathcal{Q}_n$ of non-empty sets and a \emph{protocol}, that is, a function $\Theta:\prod \mathcal{Q}_i \rightarrow \mathcal{Q}S(\rm{Im}G)$. Quantum measurement $q_{\rm{Im}G}^{meas}$ then gives a probability distribution over $\rm{Im}G$. Just as in the mixed strategy case we can then form a new game $G^{\Theta}$ by applying the expectation operator $E$. 

\begin{definition}{\rm 
Assigning the expected outcome to each probability distribution over Im$G$ that results from quantum measurement, we obtain the quantized game
$$
G^{\Theta}: \prod \mathcal{Q}_i \rightarrow \prod \Omega_i
$$
}
\end{definition}

\begin{figure}
\centerline{\includegraphics[scale=0.22]{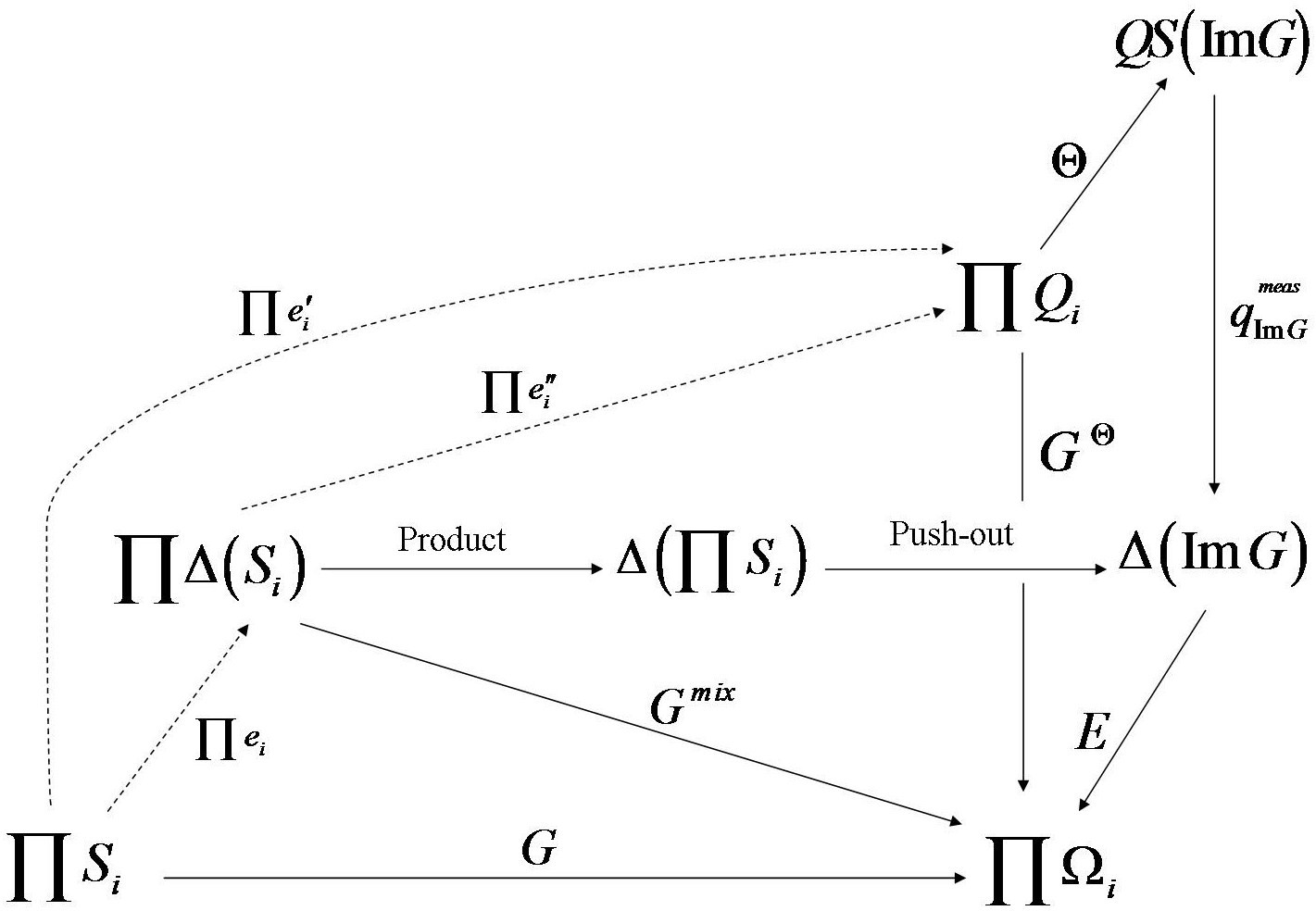}}
\caption{\small{Extension of the game $G$ to $G^{\Theta}$.}}
\label{Gquant}
\end{figure}

Call the game $G^{\Theta}$ thus defined to be the {\it quantization of $G$ by the protocol $\Theta$}. Call the $\mathcal{Q}_i$'s sets of {\it pure quantum strategies} for $G^{\Theta}$. Moreover, if there exist embeddings $e'_i:S_i \rightarrow \mathcal{Q}_i$ such that $G^{\Theta} \circ \prod e'_i=G$, call $G^{\Theta}$ a {\it proper} quantization of $G$. If there exist embeddings $e''_i:\Delta(S_i) \rightarrow \mathcal{Q}_i$ such that $G^{\Theta} \circ \prod e''_i=G^{mix}$, call $G^{\Theta}$ a {\it complete} quantization of $G$. 

This formal approach to games, and in particular game quantization, is summed up in the commutative diagram of Figure \ref{Gquant}. Note that for proper quantizations, the original game is obtained by restricting the quantization to the image of $\prod e'_i$. For general extensions, the Game Theory literature refers to this as ``recovering'' the game $G$. 

\begin{figure}
\centerline{\includegraphics[scale=0.22]{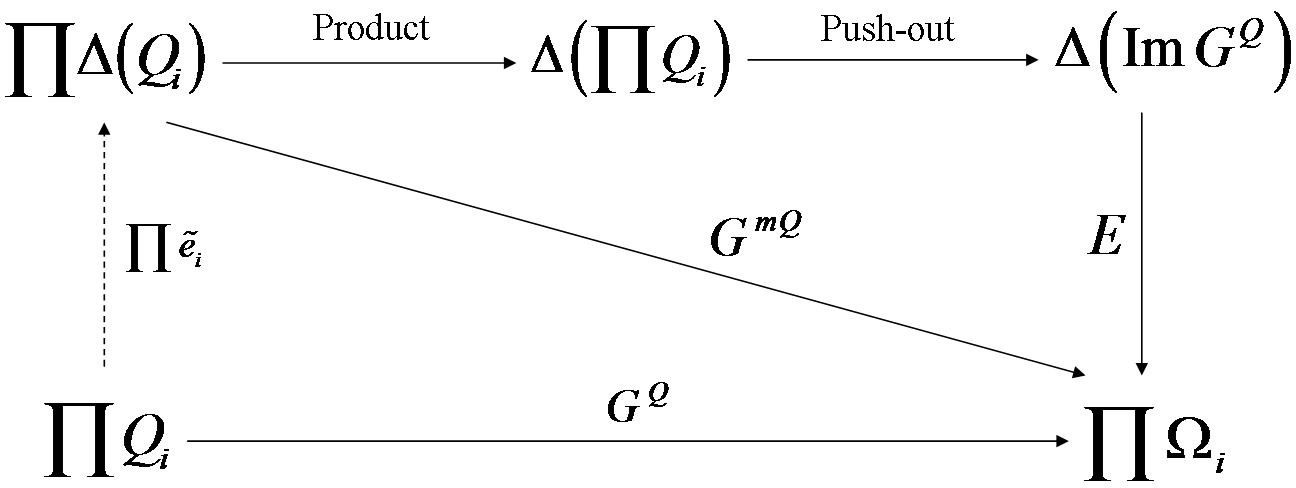}}
\caption{\small{Extension of the game $G^{\Theta}$ to $G^{m\Theta}$.}}
\label{GmQ}
\end{figure}

Though the following plays no role here,  it is worth noting that nothing prohibits us from having a quantized game $G^{\Theta}$ play the role of $G$ in the classical situation and by considering the probability distributions over the $Q_i$, create a yet larger game $G^{m\Theta}$, the {\it mixed quantization of G with respect to the protocol $\Theta$}. For a proper quantization of $G$, $G^{m\Theta}$ is an even larger extension of $G$. The game $G^{m\Theta}$ is described in the commutative diagram of Figure \ref{GmQ}.  In abstract quantum mechanics, one can access this more general notion of a mixed quantum operation directly via the consideration of completely positive operators on a quantum system, and this approach can be used to create quantum games directly.  However in this more direct construction the importance and true role of embeddings of the original and mixed games is obscured, and the existence of subgames identical to the original and mixed games becomes problematic.  This is exactly what happens in the context of Markov processes, see  \cite{Kummerer}.

\subsection{Quantizing Games with Initial States}\label{GamesWithInitial}
In many cases, the $Q_i$ of the quantization protocols are expressed as quantum operations. These operations require a state to ``operate'' on. In this situation the definition of protocol additionally requires the definition of an ``initial state'' together with the family of quantum operations which act upon this state, along with a specific definition of how these quantum operations are to act. As exemplified in the following sections, different choices for the initial state can give rise to very different protocols sharing a common selection and action of quantum operations. When a protocol $\Theta$ depends on a specific initial state $I$, the protocol is then denoted by $\Theta_I$. 

\begin{figure}
\centerline{\includegraphics[scale=0.22]{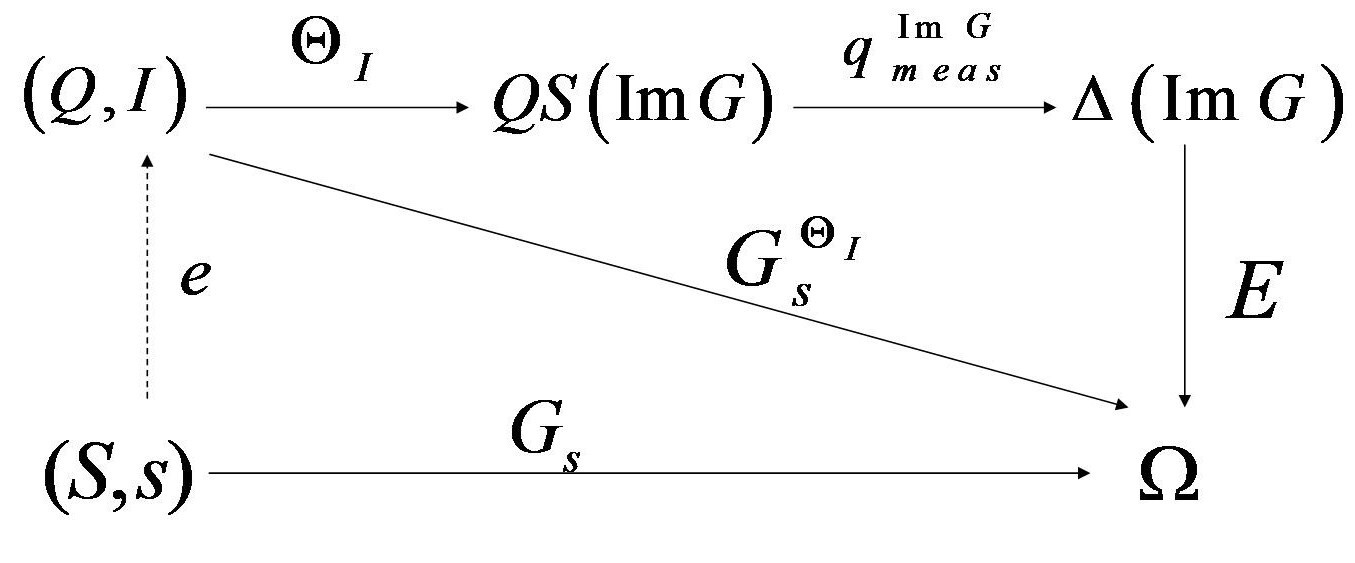}}
\caption{\small{Proper quantization of a one player game with strategy space $S$ via the protocol $\Theta$ and quantum strategy space $Q$.}}
\label{FormalParrondo}
\end{figure}

In subsequent sections, a version of the formalism adapted to one player games will be utilized to construct quantizations of history dependent Parrondo games that are in fact domain extensions. The underlying quantization paradigm being the replacement of probability distributions by the more general notion of quantum superposition followed by measurement. The functional diagram for proper quantization that will be utilized is given in Figure \ref{FormalParrondo} where the commutativity of the diagram requires that $E \circ (q^{{\rm Im}G}_{meas}) \circ \Theta \circ e =G^{\Theta} \circ e = G$. Incorporating the discussion above, when games $G_s$ and protocols $\Theta_I$ depend on a given initial states $s$ and $I$, respectively, the initial states $s$ and $I$ are regarded as part of the single player's strategic choice. In these cases, the embedding $e$ of $S$ into $Q$ additionally requires the mapping of the initial state $s$ of $G_s$ to the initial state $I$ of the protocol $\Theta_I$. The resulting quantum game is denoted by $G^{\Theta_I}_{s}$. 

\section{The FNA Quantization of History Dependent Parrondo Games}\label{QHDPG}

A major insight about quantized games that results from the formal domain extension approach to quantum games in section \ref{Bformal} is that for the quantization of a game to be game-theoretically significant, it must be proper. Previous work on the quantization of the history dependent Parrondo game by Flitney, Ng, and Abbott (FNA) \cite{Flitney:02} produced quantizations that are not proper. In this chapter, after recalling the basic facts regarding Parrondo games and the FNA quantization protocols, proper quantizations for the history dependent Parrondo game and their randomized sequences are constructed.

\begin{figure}
\centerline{\includegraphics[scale=0.25]{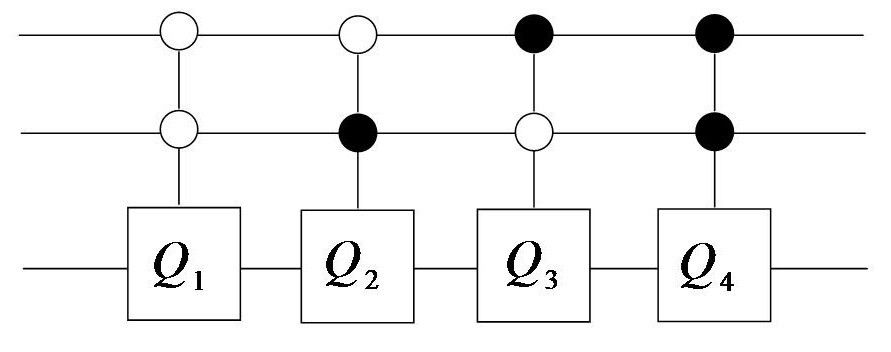}}
\caption{\small{Part of the quantization protocol for the history dependent Parrondo game. The first two wires represent the history qubits.}}
\label{HD quant}
\end{figure}

In \cite{Flitney:02}, Flitney, Ng, and Abbott quantize the type $A'$ Parrondo game by considering the action of an element of $SU(2)$ on a qubit and interpret this as ``flipping'' a biased quantum coin. They consider history dependent games with $(n-1)$ stage histories, and in the language of the Bleiler formalism, quantize these games via a family of protocols. In every protocol, $n$ qubits are required and the unitary operator representing the entire game is a $2^n \times 2^n$ block diagonal matrix with the $2 \times 2$ blocks composed of arbitrary elements of $SU(2)$. In the language of quantum logic circuits, this is a quantum multiplexer \cite{FSK:06}. The first $(n-1)$ qubits represent the history of the game via controls, as illustrated in Figure \ref{HD quant} for a two stage history dependent game similar to the game $B'$ given in Table \ref{table:gameB HD}. Each protocol is defined as the action of the quantum multiplexer on the $n$ qubits.

The quantum multiplexer illustrated in Figure \ref{HD quant}, where the elements $Q_1 \dots Q_4$ are elements of $SU(2)$, operates as follows. When the basis of the state space $(\mathbb{C}P^1)^{\otimes 3}$ of three qubits is the computational basis 
$$
\mathcal{B}=\left\{ \ket{000}, \ket{001}, \ket{010}, \ket{011}, \ket{100}, \ket{101}, \ket{110}, \ket{111} \right\}.
$$
the quantum multiplexer takes on the form of an $8 \times 8$ block diagonal matrix of the form 
\begin{equation}\label{qmuxq}
Q=\left( {{\begin{array}{*{20}c}
 {Q_1 } \hfill & 0 \hfill & 0 \hfill & 0 \hfill \\
 0 \hfill & {Q_2 } \hfill & 0 \hfill & 0 \hfill \\
 0 \hfill & 0 \hfill & {Q_3} \hfill & 0 \hfill \\
 0 \hfill & 0 \hfill & 0 \hfill & {Q_4 } \hfill \\
\end{array} }} \right),
\end{equation}
where each $Q_j \in SU(2)$. That is
\begin{equation}\label{eqn:eq16}
Q_{j}=\left( {{\begin{array}{*{20}c}
a_j \hfil & -\overline{b}_j \hfill \\
b_j \hfill & \overline{a}_j \hfill \\
\end{array}} } \right)
\end{equation}
with $a_j, b_j \in \mathbb{C}$ satisfying $\left|a_j\right|^2+\left|b_j\right|^2=1$. 

For further description of the workings of the quantum multiplexer, the following convention, found in D. Meyer's original work \cite{Meyer}, will be used. Let a ``win'' or ``gain'' for a player be represented by the action ``No Flip'' which is the identity element of $SU(2)$. For example, in Meyer's quantum penny flip game, the ``quantum coin'' is in the initial state of ``Head'' represented by $\ket{0}$ and a gain for the player using the quantum strategies occurs when the final orientation state of the coin is observed to be $\ket{0}$. This is contrast to the convention in FNA \cite{Flitney:02} where $\ket{1}$ represents a gain. 

Now the first two qubits of an element of $\mathcal{B}$ represent a history of the classical game, with $\ket{0}$ representing gain ($G$) and the $\ket{1}$ representing loss ($L$). The blocks $Q_j$ act on the third qubit in the circuit under the control of the history represented by the binary configuration of the first two qubits. For example, if the first two qubits are in the joint state $\ket{00}$, the $SU(2)$ action $Q_{1}$ is applied to the third qubit. Similarly, for the other three basic initial joint states of the first two qubits. This models the historical dependence of the game by having the history $(G,G)$ correspond to the initial joint state $\ket{00}$ of the first two qubits, the history $(G,L)$ correspond to the initial joint state $\ket{01}$, the history $(L,G)$ correspond to the initial joint state $\ket{10}$, and the history $(L,L)$ correspond to the initial joint state $\ket{11}$. Thus, an appropriate action is taken for each history. 

Recall from section \ref{HD} that the evaluation of the behavior of the classical history dependent Parrondo game requires more than just the Markov process. The evaluation also requires the stable state and a payoff rule. Note that the results of applying the quantum multiplexer depends entirely on the initial state on which it acts. That is, different initial states result in differing final states. The payoff rule used by Abbott, Flitney, and Ng resembles that for the classical game in that the quantized versions are winning when the expectation greater than $0$ (gain capital), fair if the expectation is equal to $0$ (break even), and losing if the expectation is less than $0$ (lose capital). Further, as in the classical game this question is decided by examining the probability of gain versus the probability of loss. In particular, if the probability of gain is greater than $\frac{1}{2}$, the quantum game is winning. 

\subsection{The FNA Quantization is Not Proper}

The FNA quantization protocols for the history dependent game attempt to replace the classical biases of the coins in the game with arbitrary elements of $SU(2)$ and the stable state of Markov process describing the dynamics of the game with certain initial states of the qubits on which a quantum multiplexer, composed of the arbitrary elements of $SU(2)$, acts. The problems with the FNA quantization protocols are two-fold. First, the attempted embedding of the classical history dependent game into the quantized game by replacing the biases of the classical coins with $SU(2)$ elements, turns out to be {\it relational} rather than functional. That is, Equations (\ref{qmuxq}) and (\ref{eqn:eq16}) together give a large family of quantum multiplexers that the classical game maps could be mapped into, but no restrictions on the various choices for replacement of the biased coins that could give rise to an embedding. This relational mapping makes it impossible to recover the classical game by restricting the quantized game to the image of an embedded copy of the original. Therefore, the FNA quantization of the history dependent Parrondo game is not proper. 

A second problem arises from the choice of initial state. No attempt is made in FNA to produce an analog of the stable state of the corresponding Markov process. Instead, the authors merely note that different initial states can produce different results, and in particular focus attention on two arbitrary initial states, one the maximally entangled state $\frac{1}{\sqrt{2}}\left(\left|000\right\rangle+\left|111\right\rangle\right)$, the other the basic state $\ket{000}$. In the latter, the authors assert that the quantum game behaves like a classical game with fixed initial history $(L, L)$, according to their convention in which $\ket{0}$ represents loss. Note that  this is not a proper quantization of any classical history dependent game as it fails to incorporate the other histories represented in the stable state. For 
$$
\left|000\right\rangle=
\left( {{\begin{array}{c}
1\\
0\\
0\\
0\\
0\\
0\\
0\\ 
0
\end{array}} } \right)
$$
and when acted upon by the quantum multiplexer in Equation (\ref{qmuxq}) produces the output
$$
\left( {{\begin{array}{c}
a_1\\
b_1\\
0\\
0\\
0\\
0\\
0\\ 
0
\end{array}} } \right)
$$
which makes the failure of the protocol to incorporate the other histories apparent. 

 %where the input is the initial 
%$$
%\frac{1}{\sqrt{2}}\left(\left|000\right\rangle+\left|111\right\rangle\right)
%$$
A similar situation occurs where only the histories $\ket{000}$ and $\ket{111}$ are incorporated. This protocol is also not proper as only the histories $(L,L)$ and $(G,G)$ are non-trivially represented in the initial state. For
$$
\frac{1}{\sqrt{2}}\left(\left|000\right\rangle+\left|111\right\rangle\right)=
\frac{1}{\sqrt{2}}\left( {{\begin{array}{c}
1\\
0\\
0\\
0\\
0\\
0\\
0\\ 
1
\end{array}} } \right)
$$
and when acted upon by the quantum multiplexer in Equation (\ref{qmuxq}) produces the output
$$
\frac{1}{\sqrt{2}}\left( {{\begin{array}{c}
a_1\\
b_1\\
0\\
0\\
0\\
0\\
-\overline{b_4}\\ 
\overline{a_4} 
\end{array}} } \right)
$$
from which, again, the failure of the protocol to incorporate the other histories is apparent. 

Thus, both of the FNA quantization protocols fail to reproduce the Markovian dynamics of the original history dependent Parrondo game and cannot be restricted to the payoff function of the original game.

Flitney et al also consider various ``sequences'' of the quantum games $A'$ and $B'$, where $B'$ is played with three qubits and quantized using the maximally entangled initial state. These sequences are defined by compositions of the unitary operators defining the games. Indeed, these sequences now produce the results presented in \cite{Flitney:02}. These results are certainly novel and perhaps carry scientific significance; however, they fail to carry game-theoretic significance as, with respect to the classical Parrondo games, each arises from a quantization that is {\it not} proper. 

\section{Proper Quantizations of History Dependent Parrondo Games}\label{prop quantum A and B'}

     In light of the Bleiler formalism discussed in section \ref{Bformal}, constructing proper quantizations of games is a fundamental problem for quantum theory of games. In this section, a proper quantization paradigm is developed for both history dependent Parrondo games and randomized sequences of such. 

It is crucial at this stage to view the history dependent Parrondo game discussed in section \ref{HD} in the more formal game-theoretic context of domain extension discussed in section \ref{Bformal}. For this, consider the Parrondo games as one player games as a function, where the one player's strategic choices in part correspond to the biases of the coins. For a history dependent Parrondo game with two historical stages, Parrondo et al refer to these choices as a ``choice of rules.'' However, the mere choice of biases for the coins is not enough to determine a unique normal form for these history dependent Parrondo games. In particular, an initial probability distribution over the allowable histories is also required. Although any specific distribution suffices to uniquely determine such a normal form, as the structure of the game is given by a Markov process, there is a natural choice for this initial distribution. Though this issue is not discussed by Parrondo et al, these authors immediately focus on this natural choice, namely, the distribution corresponding to the stationary state of the Markov process representing the game. 

As functions, these history dependent Parrondo games now map the tuple $\left(P, s\right)$ into the element 
$$
\left(\pi_1p_1, \pi_1(1-p_1), \pi_2p_2, \pi_2(1-p_2), \pi_3p_3, \pi_3(1-p_3), \pi_4p_4,\pi_4(1-p_4) \right)
$$ 
of the probability payoff space $[0,1]^{\times 8}$, where $s=(\pi_1,\pi_2,\pi_3,\pi_4)  \in \Delta({\rm hist}G)$ is the stationary state of the Markov process with transition matrix defined by $P=(p_1, p_2, p_3, p_4)$ $\in [0,1]^{\times 4}$, as in Equation (\ref{transition matrix}). Formally, 
\begin{equation}\label{formal payoff funct}
G_s: [0,1]^{\times 4} \times \Delta({\rm hist}G) \rightarrow [0,1]^{\times 8} 
\end{equation}
\begin{equation}\label{formal payoff funct detail}
G_s: (P,s) \mapsto \left(\pi_1p_1, \pi_1(1-p_1), \pi_2p_2, \pi_2(1-p_2), \pi_3p_3, \pi_3(1-p_3), \pi_4p_4,\pi_4(1-p_4) \right)
\end{equation}
The outcomes {\it winning}, {\it breaking even}, or {\it losing} to the player occur when $p_{\rm gain}^{B'} > \frac{1}{2}$, $p_{\rm gain}^{B'} = \frac{1}{2}$, and $p_{\rm gain}^{B'} < \frac{1}{2}$, respectively. 

Note that in this more formal game-theoretic context for history dependent Parrondo games, the dependence of these games on the initial probability distribution $s$ is made clear. This initial probability distribution plays the role of the initial state $s$ for the classical game $G_s$ appearing in the proper quantization discussion in section \ref{GamesWithInitial}. 

Consider the history dependent game $B'$ with only 2 histories. As in the FNA protocol, the quantization protocol for this game uses a three qubit quantum multiplexer with matrix representation
$$
Q=\left( {{\begin{array}{*{20}c}
 {Q_1 } \hfill & 0 \hfill & 0 \hfill & 0 \hfill \\
 0 \hfill & {Q_2 } \hfill & 0 \hfill & 0 \hfill \\
 0 \hfill & 0 \hfill & {Q_3} \hfill & 0 \hfill \\
 0 \hfill & 0 \hfill & 0 \hfill & {Q_4 } \hfill \\
\end{array} }} \right)
$$
with each $Q_j \in SU(2)$, together with an initial state.

To reproduce the classical game, first embed the four classical coins that define the game $B'$ into blocks of the matrix $Q$ corresponding to the appropriate history. The embedding is via superpositions of the embeddings of the classical actions of ``No Flip'' and ``Flip'' on the coins into $SU(2)$ given either by 
\begin{equation}\label{basic embed 1}
N=\left( {{\begin{array}{*{20}c}
1 & 0 \\
0 & 1 \\
\end{array}} } \right), \quad F=\left( {{\begin{array}{*{20}c}
0 & -\overline{\eta} \\
\eta & 0 \\
\end{array}} } \right)
\end{equation}
or by 
\begin{equation}\label{basic embed 2}
N^{*}=\left( {{\begin{array}{*{20}c}
i & 0 \\
0 & \overline{i} \\
\end{array}} } \right), \quad F^{*}=\left( {{\begin{array}{*{20}c}
0 & -\overline{i\eta} \\
i\eta & 0 \\
\end{array}} } \right)
\end{equation}
with $\eta^6=1$. Call the embeddings in equations (\ref{basic embed 1}) {\it basic embeddings of type 1} and the embedding in equations (\ref{basic embed 2}) {\it basis embeddings of type 2}. Choosing the basic embedding of type 1 embeds the $j^{\rm {th}}$ coin into $SU(2)$ as 
\begin{equation}\label{classical embed B}
Q_{j}=\sqrt{p_j}N+\sqrt{(1-p_j)}F=\left( {{\begin{array}{*{20}c}
{\sqrt p_j} & -\sqrt{1-p_j}\overline{\eta} \\
\sqrt{1-p_j}\eta & {\sqrt p_j} \\
\end{array}} } \right)
\end{equation}
where $p_j$ is the probability of gain when the $j^{\rm {th}}$ coin is played in the classical game $B'$ given in Table \ref{table:gameB HD}. Note that the probabilities $p_j$ of gaining are associated with the classical action $N$ in line with Meyer's original convention from \cite{Meyer} where $\ket{0}$ represents a gain. Hence, the elements of the subset 
$$
\mathcal{W}=\left(\ket{000}, \ket{010}, \ket{100}, \ket{110}\right)
$$ 
of $\mathcal{B}$ all represent possible gaining outcomes in the game. The probability of gain in the quantized game is therefore the sum of the coefficients of the elements of $\mathcal{W}$ that result from measurement. 

Next, set the initial state $I$ equal to 
\begin{equation}\label{unentangled state}
\frac{1}{\sqrt{\sum_{j=1}^{n}\pi_j}}\left( {{\begin{array}{c}
{\sqrt \pi_1}\\
0\\
{\sqrt \pi_2}\\
0\\
{\sqrt \pi_3}\\
0\\
{\sqrt \pi_4}\\ 
0
\end{array}} } \right),
\end{equation}
where the $\pi_j$ are the probabilities with which the histories occur in the classical game, as computed from the stationary state of the Markovian process of section \ref{HD}. The quantum multiplexer $Q$ acts on $I$ to produce the final state
\begin{equation}\label{output}
F_I=\frac{1}{\sqrt{\sum_{j=1}^{4}\pi_j}}\left( {{\begin{array}{c}
\sqrt{p_1\pi_1} \\
\eta\sqrt{(1-p_1)\pi_1}\\
\sqrt{p_2\pi_2} \\
\eta\sqrt{(1-p_2)\pi_2}\\
\sqrt{p_3\pi_3} \\
\eta\sqrt{(1-p_3)\pi_3}\\
\sqrt{p_4\pi_4} \\ 
\eta\sqrt{(1-p_4)\pi_4}
\end{array}} } \right). 
\end{equation}
Measuring the state $F_I$ in the observational basis and adding together the resulting coefficients of the elements of the set $\mathcal{W}'$ gives the probability of gain in the quantized game to be 
\begin{equation}\label{quantum prob gain}
p^{QB'}_{\rm gain}=\frac{1}{\sum_{j=1}^{4}\pi_j}\left(\sum_{j=1}^4p_j\pi_j\right)=\frac{1}{N}\left(\sum_{j=1}^4p_j\pi_j\right)
\end{equation}
which is equal to the probability of gain in the classical game. 

\begin{figure}
\centerline{\includegraphics[scale=0.23]{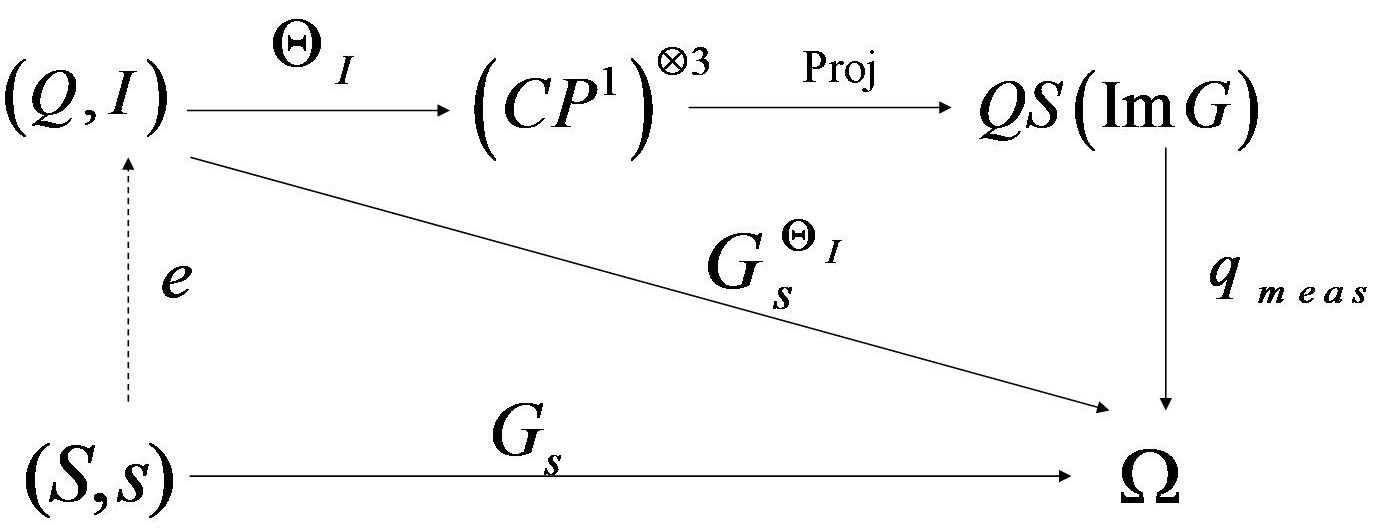}}
\caption{\small{Proper Quantization, using the embedding $e$, of the History Dependent Game via the quantization protocol $\Theta_I$}.}
\label{FormalParrondoquant}
\end{figure}

This proper quantization paradigm is based on the philosophy discussed in section \ref{GamesWithInitial}. That is, a proper quantization of a classical game $G_s$ that depends on an initial state $s$ requires that $s$ be embedded into an initial state $I$ on which the quantum multiplexer acts. Here, the initial state $s=(\pi_1, \pi_2, \pi_3, \pi_4) \in [0,1]^{\times 4}$ embeds as the initial state $I \in (\mathbb{C}P^1)^{\otimes 3}$ given in expression (\ref{unentangled state}). The resulting game $G^{\Theta_I}_s$ is the quantization of the classical game $G_s$ by the protocol $\Theta_I$ which maps the tuple $(Q,I)$, with $Q=(Q_1, Q_2, Q_3, Q_4) \in [SU(2)]^{\times 4}$ to $F_I \in (\mathbb{C}P^1)^{\otimes 3}$ given in Equation (\ref{output}). Formally,
\begin{equation}\label{quantum payoff}
\Theta_I: [SU(2)]^{\times 4} \times (\mathbb{C}P^1)^{\otimes 3} \rightarrow (\mathbb{C}P^1)^{\otimes 3}
\end{equation}
\begin{equation}\label{quantum payoff detail}
\Theta_I: (Q,I) \mapsto F_I
\end{equation}
By projecting on to the gaining basis $\mathcal{W}$, one now gets a quantum superposition over the image Im$G$ of the game $G$. Finally, quantum measurement produces Im$G$. Call $Proj$ the function that projects $F_I$ on to $\mathcal{W}$, and denote quantum measurement by $q_{meas}$. Then 
\begin{equation}\label{quantum payoff comp}
G_s^{\Theta_I}= q_{meas} \circ Proj \circ \Theta_I: (Q,I) \mapsto {\rm Im}G
\end{equation}
is a proper quantization of the payoff function of the normal form of classical history dependent game $G_s$ given in Equations (\ref{formal payoff funct}) and (\ref{formal payoff funct detail}). Equation (\ref{quantum payoff comp}) can be expressed by the commutative diagram of Figure \ref{FormalParrondoquant}, which the reader is urged to compare with Figure \ref{FormalParrondo} in section \ref{GamesWithInitial}.

Note that by embedding $s$ into $I$, the notion of randomization via probability distributions is generalized in the quantum game to the higher order  notion of randomization via quantum superpositions plus measurement. In particular, the probability distribution $P=(p_1, p_2, p_3, p_4) \in [0,1]^{\times 4}$ that defines the Markov process associated with the game is replaced with the quantum multiplexer $Q=(Q_1, Q_2, Q_3, Q_4) \in [SU(2)]^{\times 4}$ associated with the quantized game, and the stable state $s$ of the Markov process is replaced with an initial evaluative state $I$ of the quantum multiplexer.

\subsection{Proper Quantization of Randomized Sequences of History Dependent Parrondo Games}\label{sequences}

Recall from section \ref{rand A and B'} that randomized sequences of games $B'$ and $B''$ are analyzed via a Markov process with transition matrix equal to a real convex combination of the transition matrices of each game in which $B'$ is played with probability $r$ and $B''$ with probability $(1-r)$. Moreover, such a sequence is considered to by an instance of a history dependent game denoted as $B'B''$. 

Motivated by the discussion on proper quantization of the game Parrondo games $B'$ and $B''$ above, let us now consider a higher order randomization in the form of a quantum superposition of the quantum multiplexers used in the proper quantization of the the games $B'$ and $B''$ with the goal of producing a proper quantization of the game $B'B''$. 

As in section \ref{prop quantum A and B'}, associate the quantum multiplexer $Q'=(Q'_1,Q'_2,Q'_3,Q'_4)$ with the game $B'$, where
$$
Q'_{j}=\sqrt{\alpha_j}N+\sqrt{(1-\alpha_j)}F=\left( {{\begin{array}{*{20}c}
{\sqrt \alpha_j} & -\sqrt{1-\alpha_j}\overline{\eta} \\
\sqrt{1-\alpha_j}\eta & {\sqrt \alpha_j} \\
\end{array}} } \right),
$$
Next, associate the quantum multiplexer $Q''=(Q''_1, Q''_2, Q''_3, Q''_4)$ with the game $B''$, where 
$$
Q''_j = \sqrt{\beta_j}N^{*}+\sqrt{(1-\beta_j)}F^{*}=\left( {{\begin{array}{*{20}c}
{\sqrt \beta_j}i & -\sqrt{1-\beta_j}(\overline{i\eta}) \\
\sqrt{1-\beta_j}i\eta & {\sqrt \beta_j}\overline{i} \\
\end{array}} } \right).
$$
Now consider the quantum superposition
\begin{align}\label{mult superposition}
\Sigma&=\gamma'Q'+\gamma''Q''\\
&=\left( {{\begin{array}{*{20}c}
 {\gamma'Q'_1+\gamma''Q''_1} \hfill & 0 \hfill & 0 \hfill & 0 \hfill \\
 0 \hfill & {\gamma'Q'_2+\gamma''Q''_2} \hfill & 0 \hfill & 0 \hfill \\
 0 \hfill & 0 \hfill & {\gamma'Q'_3+\gamma''Q''_3} \hfill & 0 \hfill \\
 0 \hfill & 0 \hfill & 0 \hfill & {\gamma'Q'_4+\gamma''Q''_4} \hfill \\
\end{array} }} \right)
\end{align}
of the quantum multiplexers $Q'$ and $Q''$ with 
\begin{equation}\label{general quant game cond}
(\gamma')^2+(\gamma'')^2=1, \quad \left|\gamma'\right|^2=r, \quad \left|\gamma''\right|^2=(1-r), \quad \overline{\gamma'}\gamma''-\overline{\gamma''}\gamma'=0
\end{equation}
and
 \begin{equation}
\gamma'Q'_j+\gamma''Q''_j=\left( {{\begin{array}{*{20}c}
\gamma'{\sqrt \alpha_j}+\gamma''\sqrt{\beta_j}i & -\left(\gamma'\sqrt{1-\alpha_j}-\gamma''\sqrt{1-\beta_j}i\right)\overline{\eta} \\
\left(\gamma'\sqrt{1-\alpha_j}+\gamma''\sqrt{1-\beta_j}i\right)\eta & \gamma'{\sqrt \alpha_j}-\gamma''\sqrt{\beta_j}i \\
\end{array}} } \right)\\ \\
%&= \left( {{\begin{array}{*{20}c}
%\sqrt{p_j}+{\sqrt p}i & -\left(\sqrt{1-p_j}-\sqrt{1-p}i\right)\overline{\eta} \\
%\left(\sqrt{1-p_j}+\sqrt{1-p}i\right)\eta & \sqrt{p_j}-{\sqrt p}i \\
%\end{array}} } \right).
\end{equation}
Set the evaluative initial state in this case equal to 
\begin{equation}\label{AB' initial state}
I=\frac{1}{\sqrt{\sum_{j=1}^{n}\tau_j}}\left( {{\begin{array}{c}
{\sqrt \tau_1}\\
0\\
{\sqrt \tau_2}\\
0\\
{\sqrt \tau_3}\\
0\\
{\sqrt \tau_4}\\ 
0
\end{array}} } \right)
\end{equation}
where the $\tau_j$ are the probabilities that form the stationary state of the classical game $B'B''$ given in Equation (\ref{stat state mix general}). The claim is that the quantum multiplexer $\Sigma$ in Equation (\ref{mult superposition}) together with the evaluative initial state $I$ in Equation (\ref{AB' initial state}) define a proper quantization of the classical game $B'B''$ in which $B'$ is played with probability $r$ and and $B''$ is played with probability $(1-r)$. 

To check the validity of this claim, compute the output of $\Sigma$ for the evaluative initial state $I$ in Equation (\ref{AB' initial state}):
$$
\frac{1}{\sqrt{\sum_{j=1}^{n}\tau_j}}\left( {{\begin{array}{c}
{\sqrt \tau_1}(\gamma'{\sqrt \alpha_1}+\gamma''\sqrt{\beta_1}i) \\
{\sqrt \tau_1}\left(\gamma'\sqrt{1-\alpha_1}+\gamma''\sqrt{1-\beta_1}i\right)\eta \\
{\sqrt \tau_2}(\gamma'{\sqrt \alpha_2}+\gamma''\sqrt{\beta_2}i)\\
{\sqrt \tau_2}\left(\gamma'\sqrt{1-\alpha_2}+\gamma''\sqrt{1-\beta_2}i\right)\eta\\
{\sqrt \tau_3}(\gamma'{\sqrt \alpha_3}+\gamma''\sqrt{\beta_3}i)\\
{\sqrt \tau_3}\left(\gamma'\sqrt{1-\alpha_3}+\gamma''\sqrt{1-\beta_3}i\right)\eta \\
{\sqrt \tau_4}(\gamma'{\sqrt \alpha_4}+\gamma''\sqrt{\beta_4}i) \\ 
{\sqrt \tau_4}\left(\gamma'\sqrt{1-\alpha_4}+\gamma''\sqrt{1-\beta_4}i\right)\eta
\end{array}} } \right).
$$
The probability of gain produced upon measurement of this output is 
\begin{equation}\label{norm form of quant sequence}
p^{QB'B''}_{\rm gain}=\frac{1}{\sum_{j=1}^{n}\tau_j}\sum_{j=1}^4\left|{\sqrt \tau_j}(\gamma'{\sqrt \alpha_j}+\gamma''\sqrt{\beta_j}i)\right|^2
\end{equation}
which simplifies to 
\begin{equation}
\frac{1}{R}\sum_{j=1}^4\tau_j\left[\left|\gamma'\right|^2\alpha_j+\left|\gamma''\right|^2\beta_j+{\sqrt \alpha_j\beta_j}i\left(\overline{\gamma'}\gamma''-\overline{\gamma''}\gamma'\right)\right].
\end{equation}
Using the conditions set up in Equation (\ref{general quant game cond}), the previous expression further simplifies to give
$$
p^{QB'B''}_{\rm gain}=\frac{1}{R}\sum_{j=1}^4\tau_j\left[r\alpha_j+(1-r)\beta_j\right]=\frac{1}{R}\sum_{j=1}^4\tau_jt_j.
$$
which is exactly that given in Equation (\ref{p of gain in B'B''}) in section \ref{sequences} for the classical game $B'B''$.

Again, note that this proper quantization paradigm requires mapping of the initial state of the classical game $B'B''$, which is a probability distribution, into an initial state which the quantization protocol acts on, which is a higher order randomization in the form of a quantum superposition which measures appropriately with respect to the observational basis. The image of the normal form of the quantum game in $[0,1]$ agrees precisely with $p^{QB'B''}_{\rm gain}$. Note that in this proper quantization of $B'B''$, not only is the initial state of the classical game replaced by a quantum superposition, but also the probabilistic combination of the transition matrices of the classical games is replaced with a quantum superposition of the quantum multiplexers associated with each classical game. 

\subsection{A Special Case}

Recall from section \ref{rand A and B'} the classical analysis of the special case of the randomized sequence of history dependent Parrondo games, with $r=(1-r)=\frac{1}{2}$, in which one of the games is $A'$. The game $A'$ has the property that regardless of history, game $A$ is always played. Such a sequence was considered to by an instance of a history dependent game denoted by $A'B'$. In this section, a proper quantization of the randomized sequence is shown to follow as a special case of the proper quantization of the classical game $B'B''$ developed in section \ref{sequences} above.

As before, associate the quantum multiplexer $Q'=(Q'_1,Q'_2,Q'_3,Q'_4)$, where
$$
Q'_{j}=\sqrt{p_j}N+\sqrt{(1-p_j)}F=\left( {{\begin{array}{*{20}c}
{\sqrt p_j} & -\sqrt{1-p_j}\overline{\eta} \\
\sqrt{1-p_j}\eta & {\sqrt p_j} \\
\end{array}} } \right),
$$
with the game $B'$. Now, first embed the game $A$ into $SU(2)$ using basic embeddings of type 2. That is,
$$
A = \sqrt{p}N^{*}+\sqrt{(1-p)}F^{*}=\left( {{\begin{array}{*{20}c}
{\sqrt p}i & -\sqrt{1-p}(\overline{i\eta}) \\
\sqrt{1-p}i\eta & {\sqrt p}\overline{i} \\
\end{array}} } \right).
$$
The transition matrix for the game $A'$ was given in Equation (\ref{tran mat A'}) and is reproduced here:
$$
\Delta=\left(\begin{array}{cccc}
p & 0 & p & 0 \\  1-p & 0 & 1-p & 0 \\ 0 & p& 0 & p\\ 0 & 1-p& 0 & 1-p
\end{array}\right).
$$ 
The form of $\Delta$ suggests that the quantum multiplexer $Q''=(A,A,A,A)$ should be associated with the game $A'$. Now let $\gamma'=\gamma''=\frac{1}{\sqrt 2}$in Equation (\ref{mult superposition}) so that
\begin{equation}\label{mult superposition special case}
\Sigma=\frac{1}{\sqrt 2}(\Delta'+Q')=\frac{1}{\sqrt 2}\left( {{\begin{array}{*{20}c}
 {A+Q'_1} \hfill & 0 \hfill & 0 \hfill & 0 \hfill \\
 0 \hfill & {A+Q'_2} \hfill & 0 \hfill & 0 \hfill \\
 0 \hfill & 0 \hfill & {A+Q'_3} \hfill & 0 \hfill \\
 0 \hfill & 0 \hfill & 0 \hfill & {A+Q'_4} \hfill \\
\end{array} }} \right)
\end{equation}
with 
\begin{align*}
A+Q'_j&=\left( {{\begin{array}{*{20}c}
{\sqrt p}i+\sqrt{p_j} & -\left(\sqrt{1-p}(\overline{i\eta})+\sqrt{1-p_j}\overline{\eta}\right) \\
\sqrt{1-p}i\eta+\sqrt{1-p_j}\eta & {\sqrt p}\overline{i}+\sqrt{p_j} \\
\end{array}} } \right)\\ \\
&= \left( {{\begin{array}{*{20}c}
\sqrt{p_j}+{\sqrt p}i & -\left(\sqrt{1-p_j}-\sqrt{1-p}i\right)\overline{\eta} \\
\left(\sqrt{1-p_j}+\sqrt{1-p}i\right)\eta & \sqrt{p_j}-{\sqrt p}i \\
\end{array}} } \right).
\end{align*}
With the evaluative initial state 
\begin{equation}\label{AB' initial state}
I=\frac{1}{\sqrt{\sum_{j=1}^{n}\rho_j}}\left( {{\begin{array}{c}
{\sqrt \rho_1}\\
0\\
{\sqrt \rho_2}\\
0\\
{\sqrt \rho_3}\\
0\\
{\sqrt \rho_4}\\ 
0
\end{array}} } \right)
\end{equation}
where the $\rho_j$ are the probabilities that form the stationary state of the classical game $A'B'$ given in Equation (\ref{stat state mix}), the quantum multiplexer $\Sigma$ in Equation (\ref{mult superposition}) defines a proper quantization of the classical game $AB'$ when both $A$ and $B'$ are played with equal probability. 

To see this, compute the output of $\Sigma$ for the evaluative initial state $I$ in Equation (\ref{AB' initial state}):
$$
\frac{1}{\sqrt{2\sum_{j=1}^{n}\rho_j}}\left( {{\begin{array}{c}
{\sqrt \rho_1}({\sqrt p}i+\sqrt{p_1}) \\
{\sqrt \rho_1}\left(\sqrt{1-p_1}+\sqrt{1-p}i\right)\eta \\
{\sqrt \rho_2}({\sqrt p}i+\sqrt{p_2}) \\
{\sqrt \rho_2}\left(\sqrt{1-p_2}+\sqrt{1-p}i\right)\eta \\
{\sqrt \rho_3}({\sqrt p}i+\sqrt{p_3}) \\
{\sqrt \rho_3}\left(\sqrt{1-p_3}+\sqrt{1-p}i\right)\eta \\
{\sqrt \rho_4}({\sqrt p}i+\sqrt{p_4}) \\ 
{\sqrt \rho_4}\left(\sqrt{1-p_4}+\sqrt{1-p}i\right)\eta 
\end{array}} } \right).
$$
The probability of gain produced upon measurement is 
\begin{equation}\label{norm form of quant sequence}
p^Q_{\rm gain}=\frac{1}{2\sum_{j=1}^{n}\rho_j}\sum_{j=1}^4\left|{\sqrt \rho_j}({\sqrt p}i+\sqrt{p_j})\right|^2=\frac{1}{M}\sum_{j=1}^4\rho_j\left(\frac{p+p_j}{2}\right)=\frac{1}{M}\sum_{j=1}^4\rho_jq_j
\end{equation}
which is exactly that given in equation (\ref{p of win in AB'}) in section \ref{rand A and B'} for the classical game $A'B'$.
%Again, note that this proper quantization paradigm requires mapping of the initial state of the classical game $A'B'$, which is a probability distribution, into an initial state which the quantization protocol acts on, which is a higher order randomization in the form of a quantum superposition which measures appropriately with respect to the observational basis. The image of the normal form of the quantum game in $[0,1]$ agrees precisely with the expression given in equation \ref{norm form of quant sequence}. Note that in this proper quantization of $AB'$, not only is the initial state of the classical game replaced by a quantum superposition, but also a probabilistic combination of the transition matrices of the classical games $A$  is replaced with a quantum superposition of the quantum multiplexers associated with each classical game. 
 
\subsection{A Second Proper Quantization of the Randomized Sequence of History Dependent Parrondo Games}\label{second}
A second proper quantization of the sequence $B'B''$ can be constructed in a manner similar to that used to construct the proper quantization for $B'$ in section \ref{prop quantum A and B'}. Instead of forming a quantum superposition of the quantum multiplexers associated with each game, first embed the classical coins used in the game $B'B''$ into $SU(2)$ as  
\begin{align*}
Y_j &= \sqrt{t_j}N+\sqrt{1-t_j}F \\
&=\left( {{\begin{array}{*{20}c}
\sqrt{t_j} & -\sqrt{1-t_j}\overline{\eta} \\
\sqrt{1-t_j}\eta & \sqrt{t_j} \\
\end{array}} } \right)
\end{align*}
with
$$
t_j=r\alpha_j+(1-r)\beta_j \quad {\rm and} \quad 1-t_j=r(1-\alpha_j)+(1-r)(1-\beta_j)
$$ 
and associate the quantum multiplexer $Y=(Y_1, Y_2, Y_3, Y_4)$ with the classical game $B'B''$. Set the initial state, as in section \ref{sequences}, equal to
$$
I=\frac{1}{\sqrt{\sum_{j=1}^{n}\tau_j}}\left( {{\begin{array}{c}
{\sqrt \tau_1}\\
0\\
{\sqrt \tau_2}\\
0\\
{\sqrt \tau_3}\\
0\\
{\sqrt \tau_4}\\ 
0
\end{array}} } \right)
$$

\begin{figure}[t]
\centerline{\includegraphics[scale=0.26]{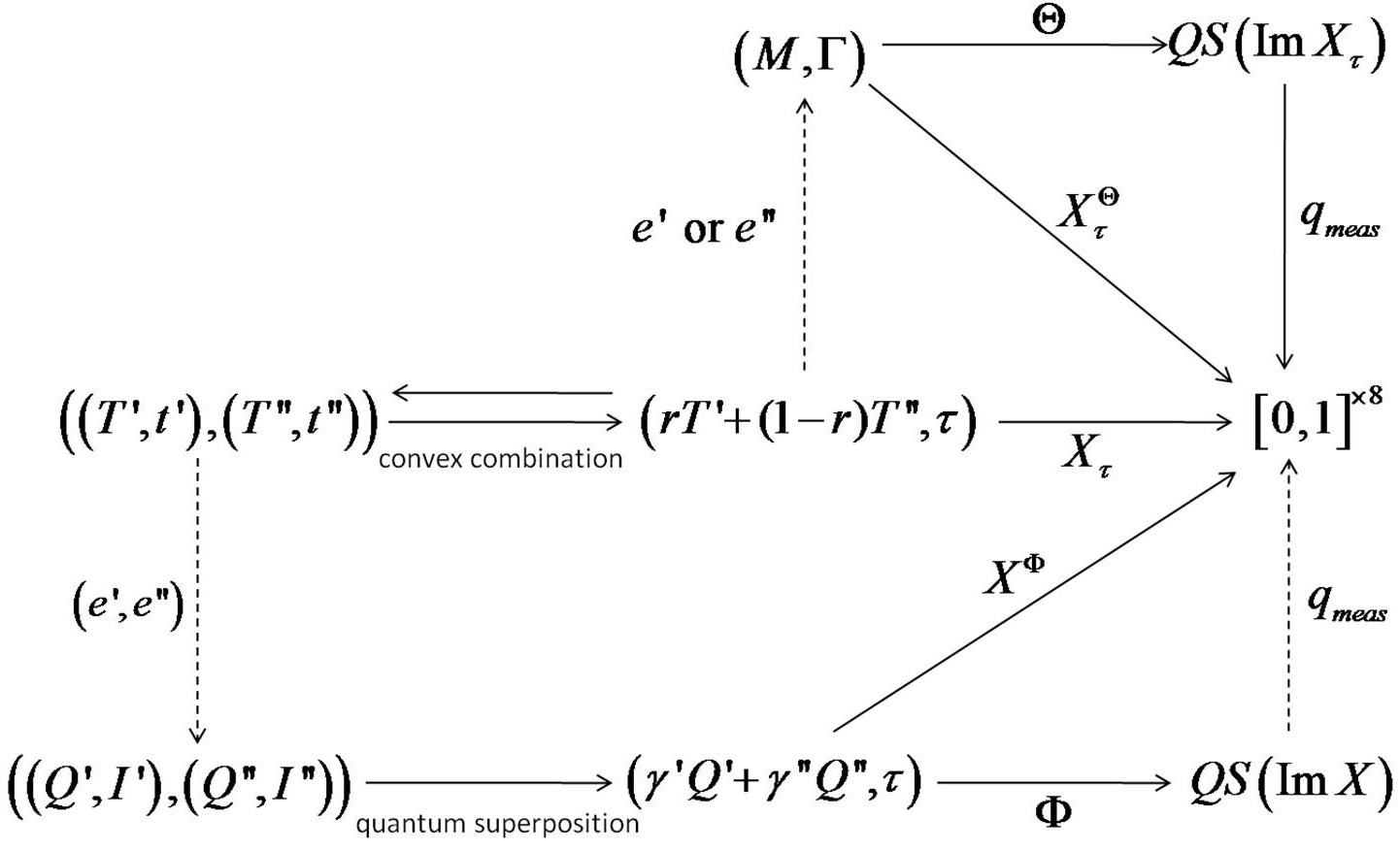}}
\caption{\small{Proper quantization of history dependent Parrondo games and their randomized sequences.}}
\label{propquantofParrondo}
\end{figure}

where the $\tau_j$ are the probabilities that form the stationary state of the classical game $B'B''$ given in Equation (\ref{stat state mix general}). The output state of this protocol is 
\begin{equation}\label{AB' output state}
F_I=\frac{1}{\sqrt{\sum_{j=1}^{n}\tau_j}}\left( {{\begin{array}{c}
\sqrt{\tau_1t_1}\\
\sqrt{\tau_1(1-t_1)}\eta\\
\sqrt{\tau_2t_2}\\
\sqrt{\tau_2(1-t_2)}\eta\\
\sqrt{\tau_3t_3}\\
\sqrt{\tau_3(1-t_3)}\eta\\
\sqrt{\tau_4t_4}\\ 
\sqrt{\tau_4(1-t_4)}\eta
\end{array}} } \right)
\end{equation}
which, upon measurement produces the probability of gain
$$
p^{QB'B''}_{\rm gain}=\frac{1}{\sum_{j=1}^{n}\tau_j}\sum_{j=1}^4\tau_jt_j
$$
which is exactly the probability of gain computed in Equation (\ref{p of win in AB'}) of section \ref{rand A and B'} for the classical game $AB'$.

\section{Conclusions}
Two approaches are used to properly quantize random sequences of Parrondo games $A$ and $B'$ in which each game occurs with equal probability. One approach, discussed in section \ref{prop quantum A and B'}, generalizes the notion of randomization between the two games via probability distributions to randomization between games via quantum superpositions. The other approach, discussed in section \ref{second}, embeds a probabilistic combination of the games into a quantum multiplexer directly rather than via quantum superpositions of the protocols for each game. In the former approach, note that it was crucial that game $A$ was embedded into $SU(2)$ using basic embedding of type 2 as this allowed for the use of the broader arithmetical properties, namely factorization, of complex numbers to reproduce the classical result. In the latter on the other hand, basic embedding of type 1 sufficed. 

\section{Future Directions}

The ideas developed in this article bring together formal game theory, Markov processes, and quantum information theory. Due to this multifaceted nature, the study of proper quantization of games can potentially influence research in all three areas mentioned above. For instance, the proper quantization protocols developed for history dependent Parrondo games using a particular type of quantum multiplexer lend a game theoretic perspective to the study of quantum logic circuits via quantum multiplexers. Indeed, the notion of the Parrondo effect is now attached to quantum circuits and it is now natural to investigate the characterization of the ``quantum Parrondo effect'' in quantum circuits via a game theoretic perspective. 

Results in quantum logic synthesis show that an $n$ qudit logic gate can be synthesized via a circuit consisting entirely of variations of the quantum multiplexer \cite{FSK:06, Brylinski:02}. Given the interplay of game theory and quantum circuits in the quantization of history dependent Parrondo games, it is also natural to ask how might an arbitrary quantum logic gate be synthesized via a quantum multiplexer circuit in a game theoretically meaningful way.  For example, after assigning a fixed number of qubits in the circuit to each "player", for an arbitrary quantum logic gate $U$, how might $U$ be decomposed into a quantum multiplexer circuit and an initial state chosen such that a given game theoretic outcome might be realized?

In an even broader context, to date there is no agreement in the literature on exactly what a quantum Markov process is. One difficulty lies with the formulation of an appropriate definition of the ``quantum'' analogue for the stable state of a classical process, an object here called the {\it evaluative} state. Our quantizations of history dependent Parrondo games are specially quantized Markov processes involving specific elements of the Lie group $SU(2)$ and with evaluative states chosen game-theoretically. A more general situation exists in which arbitrary elements of $SU(2)$ are utilized. In such a situation, one asks if it is possible to use quantum game theory to come up with a natural choice for the evaluative state.  Moreover, one also asks if it is possible to characterize a quantized version of the Parrondo effect in this general set up, and if so, what does such a characterization mean for quantum computation? 

%---------------Bibliography-------------------------------
\singlespace

%\bibliography{BKrefs}
%\bibliographystyle{plain} 

\end{document}